\documentclass[preprint2]{aastex}
\begin{document}
\title{Spectroscopic classification of cataclysmic variable candidates
\thanks{Based on observations collected at the 
        NOAO Cerro Tololo Interamerican Observatory, Chile 
        and the European Southern Observatory, La Silla, Chile. }}

\author{L. Schmidtobreick}
\affil{European Southern Observatory, 
          Casilla 19001, Santiago 19, Chile }
\email{lschmidt@eso.org}
\author{L. Galli}
\affil{Colorado College, 14 East Cache La Poudre St., 
          Colorado Springs, CO 80903 USA.}
\author{A. Whiting}
\affil{Cerro Tololo Inter-American Observatory, 
          Casilla 603, La Serena, Chile}

\author{C. Tappert}
\affil{Departamento de Astronom\'{\i}a y Astrof\'{\i}sica,
           Pontificia Universidad Cat\'olica, Casilla 306, Santiago 22, Chile}
\and
\author{A.J. Carver}
\affil{University of Wisconsin, Madison, USA}


\begin{abstract}
We present low resolution optical spectroscopy for six cataclysmic variable 
candidates and the old nova V888\,Cen. 
We confirm the classification as cataclysmic variable
for LB\,9963 and FQ\,Mon, while
the other four candidates turn out to be different type of stars.
We discuss the individual spectra and pay special attention to the
mass transfer rate and disc temperature and density of the
three cataclysmic variables. 
\end{abstract}




\section{Introduction}
Cataclysmic Variables (CVs) are close, interacting binary systems,
with a white dwarf primary receiving mass
from a Roche--lobe--filling late--type star.
In absence of strong magnetic fields, the mass transfer takes 
place via an accretion disc; otherwise the matter is channeled 
along the magnetic field lines directly onto the surface of the white dwarf
\citep[see][for a thorough introduction to these objects]{warn95}.

The online edition of the CV catalog of \citet{down+01} 
contains many 
objects with uncertain classification. The majority of these do not have
any published spectra, which are essential for a classification and for
the confirmation of the CV nature of these objects.

\begin{deluxetable}{lcrlcr}
\tablewidth{0pt}
\tablecaption{\label{obstab} Summary of the observational details.}
\tablehead{
\colhead{Object} & \colhead{RA$_{2000}$} & \colhead{DEC$_{2000}$} & 
\colhead{Telescope/Instrument} & \colhead{Date} & 
\colhead{$t_{\rm Exp}$ [s]}}
\startdata
FBS\,0204-024 & 02:06:44.5 & -02:12:17 & CTIO1.5m/R-C & 2004-02-06 & $2 \times 1200$ \\
              &            &           & CTIO1.5m/R-C & 2004-02-09 & 1800 \\
LB\,9963      & 02:50:24.6 & -87:30:23 & CTIO1.5m/R-C & 2004-02-06 & $3 \times 900$ \\
              &            &     & ESO3.6m/EFOSC/Gr10 & 2004-03-16 & 600 \\
              &            &      & ESO3.6m/EFOSC/Gr7 & 2004-03-16 & 600 \\
WY CMa        & 07:11:40.1 & -26:58:40 & CTIO1.5m/R-C & 2004-02-06 & $2 \times 900$ \\
              &            &           & CTIO1.5m/R-C & 2004-02-09 & $3 \times 900$ \\
FQ Mon        & 07:16:41.2 & -06:56:49 & CTIO1.5m/R-C & 2004-02-06 & $3 \times 1800$ \\
              &            &      & ESO3.6m/EFOSC/Gr6 & 2004-03-16 & $3 \times 600$ \\
              &            &      & ESO3.6m/EFOSC/Gr6 & 2004-04-31 & $3 \times 600$ \\
              &            &      & ESO3.6m/EFOSC/Gr6 & 2004-11-14 & $3 \times 600$ \\
              &            &      & ESO3.6m/EFOSC/Gr6 & 2004-11-19 & $3 \times 600$ \\
FV Cnc        & 08:48:01.8 &  18:40:37 & CTIO1.5m/R-C & 2004-02-09 & $3 \times 300$ \\
V591 Cen      & 12:42:18.1 & -33:34:10 & CTIO1.5m/R-C & 2004-02-06 & $3 \times 1800$ \\
V888 Cen      & 13:02:31.9 & -60:11:36 & CTIO1.5m/R-C & 2004-02-07 & 1200 \\
              &            &           & CTIO1.5m/R-C & 2004-02-09 & $3 \times 1800$ \\
\enddata
\end{deluxetable}

As part of the REU (Research Experiences for Undergraduates) 
observation campaign at CTIO (Cerro Tololo Inter-American Observatory)
in February 2004, we have taken optical spectra of six CV--candidates
with the purpose of classification. We here present the results on these
stars. We also include the first optical spectrum of Nova 1995 Cen (V888\,Cen)
in quiescence.

\section{Data and reduction}
\label{data}

We observed the CV--candidates in low resolution 
with the  R-C spectrograph at the 1.5\,m telescope at CTIO. 
Some follow--up observations have been done with EFOSC at the 
3.6m telescope of ESO, La Silla,
see Table \ref{obstab} for the details.
The spectra were all taken with the spectrograph slit
aligned with the parallactic angle.
Standard reduction was performed for all data with IRAF.
The bias were subtracted and the data were divided by a flat field,
which was normalized by fitting Chebyshev functions of high order 
to remove the detector specific spectral response. For those objects where the
observation had been divided in several exposures, the individual frames 
have been averaged.
The spectra have been optimally extracted \citep{horn86}.
Wavelength calibration yielded a final FWHM resolution of 1.5\,nm
and a spectral range of 370--970\,nm for the CTIO data. 
The EFOSC spectra were taken with a slitwidth of 1.0".
For grism \#6 this yields a FWHM resolution of 1.3\,nm with
a range of 390--800\,nm, for grisms \#7 and \#10, we get a resolution of
0.6\,nm and a range of 330-520\,nm and 630-820\,nm respectively.
The CTIO spectra have been normalized by fitting splines to the continuum.
The La Silla spectra have been corrected for the instrument curve using
standard star observations.
While the absolute
flux values have to be regarded with caution, the relative fluxes can be
used to compare different parts of the spectrum.

For LB\,9963 we also performed time resolved V--photometry using the 90\,cm
telescope at CTIO. The reduction was done using the
quadproc package in IRAF. The data have been corrected for overscan, bias and 
flatfields. Aperture photometry for LB\,9963 and three reference stars was done
with apphot inside IRAF.

All subsequent analysis of the data has been done using MIDAS.

\section{Results}

\begin{figure*}
\includegraphics[scale=0.85]{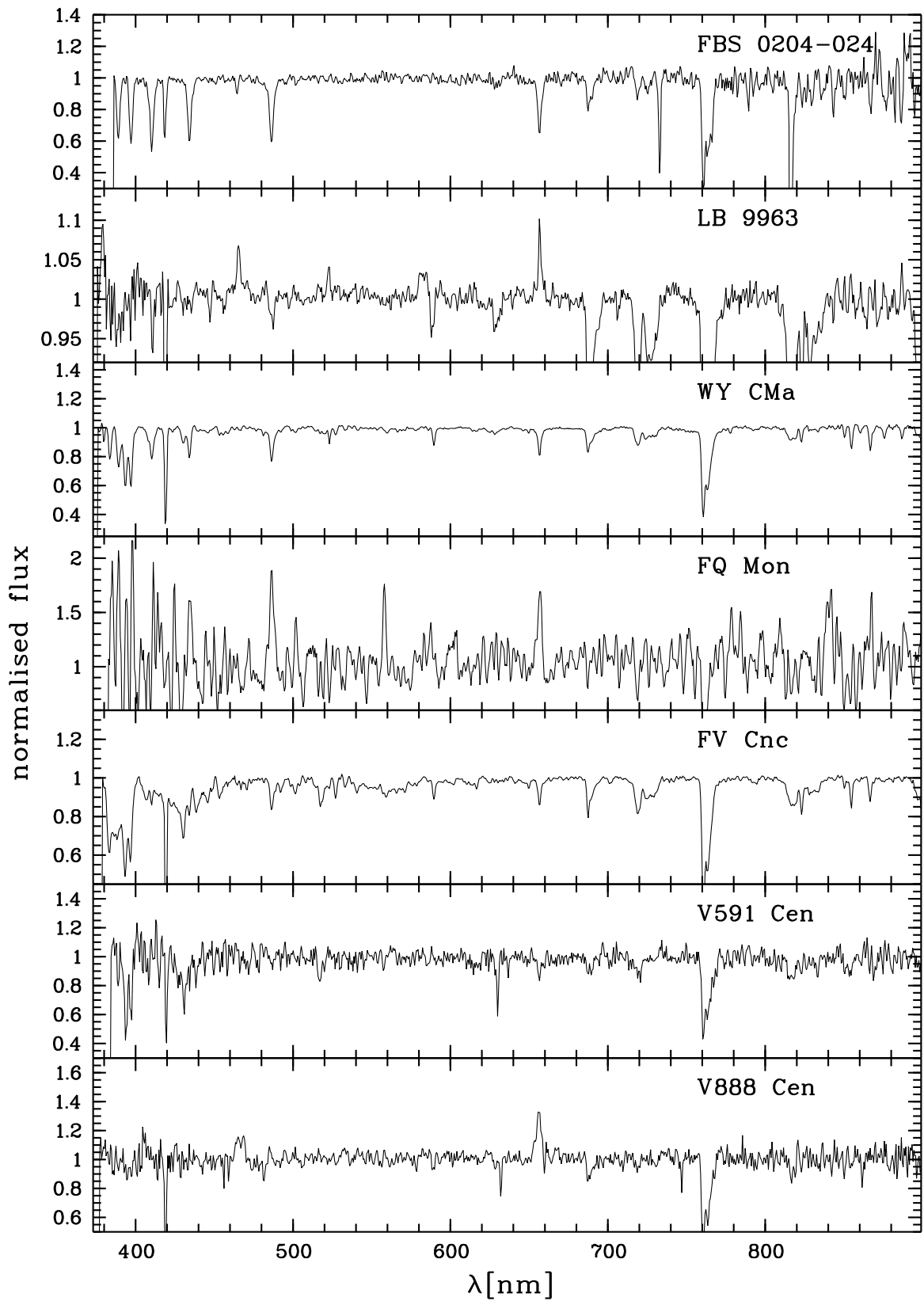}
\caption{\label{spec_all}The spectra of all objects 
observed at CTIO in the range 370--900\,nm. The continuum has been 
normalized to unity. Note that the
absorption feature at 420\,nm is a CCD artefact.}
\end{figure*}

In Fig. \ref{spec_all}, the normalized CTIO--spectra are plotted for all 
observed objects. In the following we will discuss these objects in detail.

\subsection{FBS 0204-024}

The background information on this object is rather unclear. In \citet{down+01}
it is named Cet and given as a candidate CV. For the classification they 
refer to
\citet{abra+03} where the object has the catalog number FBS 0204-024. 
However, Abrahamyan et al. classified 
this object as a B2e star by prism spectroscopy
without mentioning a possible CV classification. 
They identified the star in the catalog of \citet{berg+80}, possibly as 
object PB 6657, which is the closest coordinate match. 
Still, PB 6657 lies about 2.5 arcmin
to the south and is 3 mag fainter than FBS 0204-024, so the identification 
seems rather doubtful.
The finding chart provided by \citet{down+01} is established via 
coordinate match with FBS 0204-024. 

The spectrum, that we show in Fig.~\ref{spec_all}, is taken of this object.
It shows the object to be a late B-type star; best match of the 
absorption lines has been achieved with a B6-9V template. Note that we find
no evidence for emission lines, in contrast to the classification by 
\citet{abra+03}.
Since Be stars are known to also show phases without 
emission lines, we restrain ourselves from comments on the specific nature 
of this object.

\subsection{LB\,9963}

\begin{figure}[t]
\includegraphics[scale=0.35]{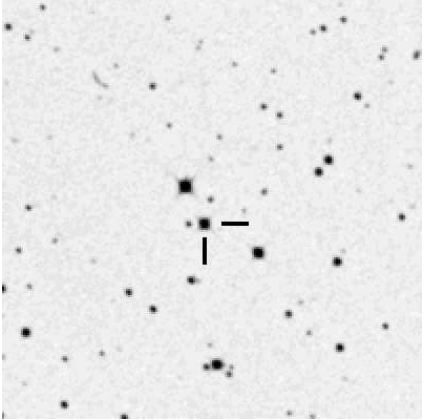}
\caption{\label{fc_oct}A $5^{\prime}\times 5^{\prime}$ chart of the
DSS is plotted, north is up, east is left. The object LB\,9963 
at RA = 02:50:22, DEC = -87:30:23} is indicated.
\end{figure}

\begin{figure}[t]
\includegraphics[angle=-90,scale=.29]{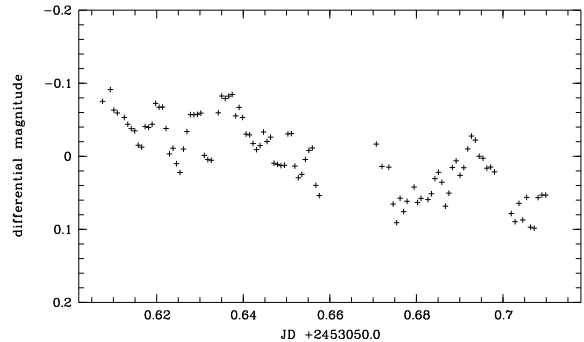}
\caption{\label{oct_lc} The differential magnitude of Oct calculated from 
three stable comparison stars are plotted against Julian date. The 
uncertainties of the differential photometry are 0.01 mag, which is 
about the size of the symbols used for the plot.}
\end{figure}

The object was first investigated by \citet{kilk95}. 
From Str\"omgren photometry he classified the object as a 
heavily reddened hot subdwarf: sdOr. He also presents a low resolution 
spectrum covering the range 3600--5000\AA. Although very noisy, faint 
Balmer emission lines can be seen, which lead to the classification as
a possible CV.

No candidate is given on the finding chart provided by 
\citet{down+01}, where the object is called Oct. 
From private communication with Dave Kilkenny we identified the 
probable candidate at RA\,=\,02:50:22, DEC\,=\,-87:30:23 (J2000)
which we indicate in the chart in Fig.~\ref{fc_oct}.
The CTIO spectrum of this object (Fig.~\ref{spec_all}) shows faint and 
narrow H$\alpha$ emission but
no other Balmer emission lines. H$\beta$ is present in absorption with 
a slight hint of an emission core. The same is found for H$\gamma$ and 
H$\delta$, although the spectrum gets very noisy in this range. 
He\,I at 588\,nm is found in absorption but blended with the Na doublet at
589\,nm. No other He lines are found in this spectrum. 
Two emission features are found at 581\,nm and 465\,nm. They are probably
due to the C\,IV doublet (580/1\,nm) and the Bowen blend of N\,III, C\,III,
and C\,IV at 465\,nm. Similar features have been found in a few other
CVs \citep[for a summary, see][]{schm+03}, usually indicating a very high mass
transfer rate and a probably evolved secondary.

\begin{figure*}[t] 
\includegraphics[angle=-90,scale=.59]{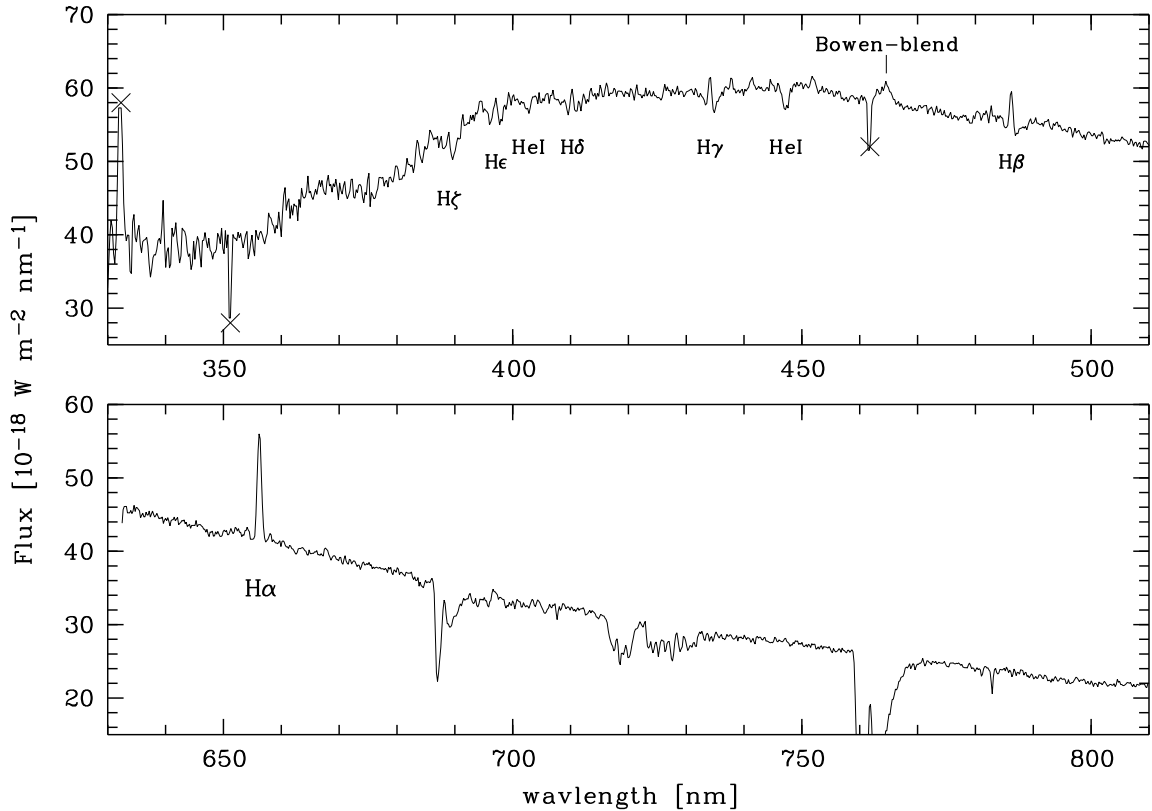}
\caption{\label{ls_oct}The spectra of LB\,9963 have been flux--calibrated with
an accuracy of 10\%. The upper plot shows the blue spectrum between 330 and 
510\,nm, the lower plot shows the red spectrum between 630 and 810\,nm.
Three artefacts in the upper plot, which are due to strong cosmics, 
are marked ($\times$).}
\end{figure*}

The classification as a cataclysmic variable or at least a similar object
is confirmed by a short photometric run. For observational details see
section \ref{data}. The lightcurve is plotted in Fig.~\ref{oct_lc}. A general
decrease of brightness might be due to a periodical variation, which would then
imply an orbital period of at least five hours. However, the more important
variation is the observed flickering of the star which is a valuable 
evidence for the CV classification of LB\,9963.

To verify the presence of emission cores in the Balmer absorption lines we 
performed follow--up observations with EFOSC2 at the 3.6m telescope at La Silla
as described in section \ref{data}.
The two spectra  (one red, one blue) have been flux--calibrated with an 
accuracy of about 10\% but not
corrected for interstellar extinction. They are plotted in Fig.~\ref{ls_oct}.
In the red spectrum, H$\alpha$ is clearly seen in emission. In the blue spectrum,
emission cores
are confirmed in the absorption troughs of H$\beta$, H$\gamma$, H$\delta$,
and  H$\epsilon$. All the blue He\,I lines are found in absorption, and only
a faint emission is found for He\,II at 468.6\,nm. We find no indication of
He\,I in the red spectrum. The strongest emission in the blue is found at
465\,nm and can be associated with the N\,III, C\,III and C\,IV emission in the
Bowen blend. From the properties of the emission lines, LB\,9963 would thus be
characterized as a novalike variable with high mass transfer and accordingly
optical thick accretion disc of high temperature. Still, the absence
or weakness of He\,II is puzzling.

Such a hot disc however, is supposed to have a strong blue continuum.
The continuum of LB\,9963, instead, shows a steep rise between 
350\,nm and 400\,nm, 
which is not at all typical for a cataclysmic variable. 
We have examined our reduction process several times and cannot find any 
indication of it artificially producing this unusual continuum.
The spectrum has been taken 
at a high airmass of 1.9 but under parallactic angle to minimize the effect 
of differential refraction. Furthermore, the spectrum of \citet{kilk95} 
shows a similar behavior. We thus believe this continuum to be real.

Since we have only little information
on the interstellar extinction, the cut towards the blue might be due to an 
exceptionally high interstellar reddening. Although we cannot rule out this
possibility, we believe it rather unlikely as the galactic latitude of 
LB\,9963 is $b =-29^{\circ}$ and the average extinction in this area of the sky
is with $A_V \approx 0.05$ rather low \citep{drim+03}. 
However, a thin dust filament is distinguished in the IRAS
maps of this region and might enhance the local extinction of LB\,9963.
The higher resolved dust maps of \citet{schl+98} e.g. yield a 
reddening $E_{B-V} = 0.14$. This value is supported by the equivalent width 
of the Na\,I absorption line. We find $W_{\rm Na} = 0.14(1)$\,nm for 
Na\,I (D1 + D2) which converts to $E_{B-V} = 1.1(2)$ using the 
empirical relation discussed by \citet{muna+97}.  
We used the equations of \citet{howa83} to deredden the 
spectra with either value.
It does not improve the continuum shape significantly.
In fact, a reddening as high as $E_{B-V} \approx 3$ is needed to get a "normal" 
novalike continuum.

Assuming that the disc and the white dwarf do not contribute a lot to the 
continuum, alternatively, a late F-type secondary could explain the shape 
of the continuum.
One would then expect to also see the absorption lines of this star, which would 
thus naturally explain the presence of the Balmer absorption troughs without 
assuming a hot disc. However, this does not explain the He absorption, which 
is not present in F-type stars.
Also, the Ca\,II--line at 393.4\,nm, which is typically a clear indication
for an F--type star, is not present in the spectrum. We thus rank this 
possibility of an F-type secondary very low.

On the balance, we think that, however unlikely, a highly reddened 
novalike variable
matches the spectrum better than a CV with F-type star secondary.

\subsection{WY Canis Majoris}
WY CMa has been classified as a $\delta$ Cepheid by \citet{hoof41} 
who also gives
the period of 1.14\,days and a photographic amplitude of 0.45$^{\rm m}$. 
Subsequently, the star appears in several 
catalogs and investigations, e.g. \citet{petit60b} or \citet{pete+87}.
\citet{hack+90} re-observed WY CMa photographically
and found that the resulting lightcurve does not agree with the classification
as a Cepheid and assume that the object is rather a long--period CV. 
Thus, the object has been included in the CV catalogs by \citet{ritt+98}
and \citet{down+01}.

Our spectrum shows the object to be an early F--type star, probably a giant.
Best matches of the absorption lines have been achieved with a F2III template.
The absence of any CV features in the spectrum make its
designation as such untenable.  Further photometry, especially with a
high-precision detector, might reopen the possibility of it being indeed a
Cepheid or allow its assignment to a different category of variable.

\subsection{FQ Monocerotis}

\begin{deluxetable}{lrrrrrrrrrrrr}
\tabletypesize{\small}
\tablewidth{0pt}
\tablecaption{\label{line_tab}
FWHM [in nm], equivalent widths [in nm], and line fluxes 
[in $10^{-18}$\,W\,m$^{-2}$] of the Balmer emission lines
are given for the four observations of FQ\,Mon in quiescence.}
\tablehead{
& \multicolumn{3}{c}{Feb 06} & \multicolumn{3}{c}{Apr 30}& \multicolumn{3}{c}{Nov 13}& \multicolumn{3}{c}{Nov 18} \\
\colhead{}& \colhead{fwhm} & \colhead{-W} & \colhead{F} & 
\colhead{fwhm} & \colhead{-W} & \colhead{F} &
\colhead{fwhm} & \colhead{-W} & \colhead{F} &
\colhead{fwhm} & \colhead{-W} & \colhead{F}}
\startdata
H$\alpha$  &3.2(2)&3.1(3)&3.0(3)&2.9(1)&0.5(1)&0.7(1)&2.9(2)&0.5(1)&0.9(2)&      &      &      \\
H$\beta$   &4.0(1)&4.5(2)&8.9(4)&3.8(3)&0.5(1)&0.8(1)&3.1(1)&1.0(1)&1.6(1)&      &      &      \\
H$\gamma$  &3.1(2)&5.9(4)&12.4(9)&3.0(3)&1.2(2)&1.3(2)&2.4(1)&2.3(1)&3.1(1)&1.7(3)&0.4(1)&0.6(1)\\
H$\delta$  &      &      &      &3.1(3)&1.5(2)&1.5(2)&2.0(1)&2.3(2)&3.0(2)&1.5(1)&0.7(1)&1.1(1)\\
H$\epsilon$&      &      &      &      &      &      &1.6(1)&2.6(1)&2.9(1)&1.7(1)&0.8(1)&1.0(1)\\
H$\zeta$   &      &      &      &      &      &      &1.7(1)&2.9(1)&2.6(1)&      &      &      \\
\enddata
\end{deluxetable}

FQ\,Mon has first been mentioned as a variable star by \citet{hoffm36}
who discovered an outburst in January 1929. 
No further analysis of the system had been done. In March 2004, about
three weeks after our observation, FQ\,Mon went into superoutburst 
with $V\approx 14.5$\,mag, thus
showing it to be a SU UMa system \citep{uemur04}. 
The final superhump period has been determined as 
1.634\,h \citep{masi04}.

Our spectrum (Fig.\ref{spec_all}) confirms the CV-nature of FQ\,Mon.
Although noisy, it is dominated by the Balmer lines in emission.
Also the Paschen series and He\,I are detected in emission. 
The spectrum resembles thus a typical dwarf nova spectrum in quiescence, 
although the inverse Balmer decrement (see Tab.~\ref{line_tab}) 
points to a higher temperature or density of the accretion disc.

\begin{figure}[t]
\includegraphics[scale=.44]{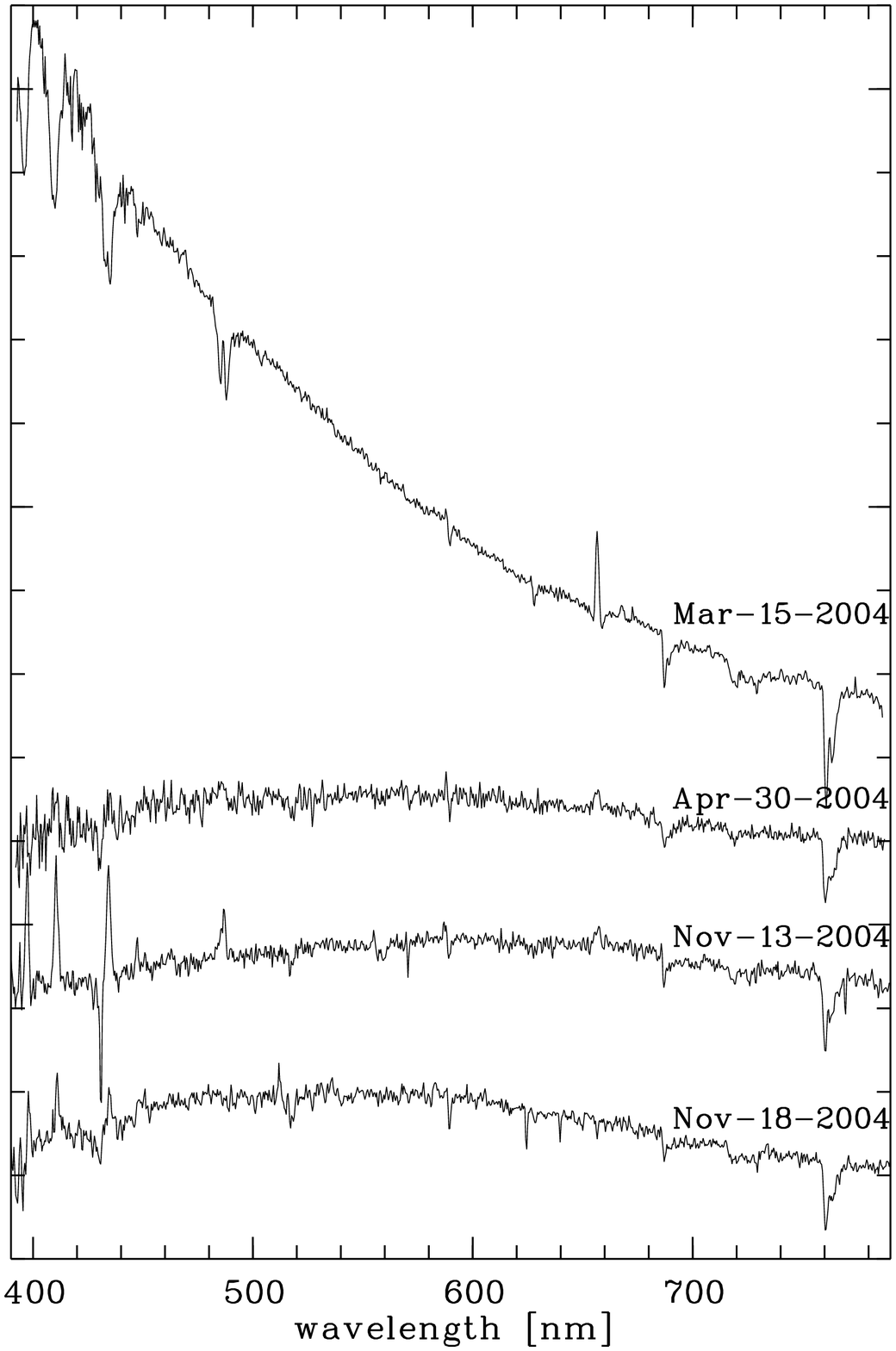}
\caption{\label{ls_fqmon} The spectra of FQ\,Mon have been corrected for the
instrument function and then arbitrarily shifted. Note that even in quiescence
a variation in the shape of the continuum is present.}
\end{figure}

Follow--up observations have been done for this object on La Silla. 
The resulting four spectra are plotted in Fig.\,\ref{ls_fqmon}. 
The spectrum from March clearly confirms that FQ\,Mon was
in outburst during this epoch. It is dominated by the very blue continuum 
and only faint emission peaks are present within absorption troughs. 
From the acquisition files we have estimated the brightness of FQ\,Mon
by comparing it to the stars U0825--04435970, U0825--04433933, and 
U0825--04434176 from the USNO-A2.0 catalog. We find $V=15.3(2)$, $V=21.7(3)$, 
$R=20.5(3)$ and $R=20.3(3)$ for Mar\,15, Apr\,30, Nov\,13, and Nov\,18.
We thus estimate the amplitude of the superoutburst to be of the order of 
7\,mag.

The spectra also show that FQ\,Mon is a highly variable object.
Even the quiescence spectra are not alike but show variations of the continuum
shape as well as of the strengths of the various emission lines
(see Tab.~\ref{line_tab}).
Note that there is a slightly brighter star very close to FQ\,Mon.
We cannot completely rule out the 
possibility, that some contribution in the spectrum comes from this close 
companion and might produce an artificial variation. However, since the
seeing always allowed us to separate the two stars and since
we took special care to place the slit on FQ\,Mon itself, 
an additional contribution appears rather unlikely.

The emission lines are clearly resolved
and yield projected rotational velocities of about 1500\,km/s, thus 
indicating a high inclination system. Still,
although broad, the lines show no double peak profile as should be expected
for a high inclination dwarf nova. Furthermore, even in the spectrum with the
strongest emission lines, H$\alpha$ is rather weak, even weaker than in the 
spectrum taken during outburst, and the Balmer lines show 
a negative decrement. This is usually an indication for a magnetic system,
so FQ\,Mon might thus belong to this category of CVs. 
Since the superoutburst clearly classifies it as a SU\,UMa system, the
magnetic field can only be of mediocre strength. In this respect FQ\,Mon might be
comparable to GZ\,Cnc, a dwarf nova at the lower edge of the period gap, which
is supposed to be a magnetic system and also shows an inverse 
Balmer decrement, although less strong than is the case for FQ\,Mon
\citep{tapp+03}. Another similar system is VZ\,Pyx, also a SU\,UMa system 
which is at the same time a candidate intermediate polar. 
Again, a slightly inverse Balmer
decrement is found in the spectrum \citep{remi+94}. Note that FQ\,Mon can
be identified with the ROSAT source RXS\,J071641.6-065653, which 
could also indicate a magnetic nature.

\subsection{FV Cancri}
FV\,Cnc has been classified as a possible CV of U\,Gem subtype by
\citet{kaza+99} due to an $0.4^{\rm m}$ intensity increase observed by Hipparcos.
The object was known as member of the open cluster Praesepe
on grounds of photometry and proper motion (\citet{merm+90}; \citet{jone+91}).
Mermilliod et al. also found that FV\,Cnc is a double--lined
spectroscopic binary with
an orbital period of 2.981781(7) days and a circular orbit. They present
the velocity curves of both components and derive a mass ratio of 1.06. 

Our spectrum shows the object to be a G-type star; best match of the 
absorption lines has been achieved with a G0-4 template. 
Due to the low resolution of our data,
we cannot resolve the double--line character.
The classification is confirmed by the published photometric colors.
Mermilliod et al. measured (B-V) = 0.69 and (U-B) = 0.19, the colors of
a G5V star, while (B-V) = 0.572 is given in the Tycho catalog, which thus 
favors a slightly earlier type.

The spectrum of FV Cnc is thus inconsistent with a CV designation, and shows 
it to be a close binary consisting of two early G-type stars.  The 
brightness variation observed by HIPPARCOS remains to be explained. It 
could have its source in magnetic activity, or some external (third-body) 
cause.

\subsection{V591 Centauri}
The object has first been noted as a variable by \citet{petit60a}, 
who included it as a possible long period star in his 
catalog of U\,Gem like stars. However, in the supplements to this catalog 
\citep{petit61}, he marks the variability as irregular.

Our spectrum is rather noisy but resembles a normal G-type star spectrum.
A best match is achieved for a G5-8V template. Hence, the variability is 
probably due to magnetic activity.

\subsection{V888 Centauri}

The nova V888\,Cen was discovered by Liller on February 23, 1995 at a visual 
magnitude of 7.2 \citep{lille95}. \citet{yant+01} have analyzed the outburst
spectra and the lightcurve. They found the nova to be very fast with 
$t_2$ = 5 days and to show strong oscillations in the transition region.
The spectra were dominated by strong emission lines with P Cygni profiles
yielding velocities of -1765 and -3010 km/s. The presence of 
Fe\,II emission lines puts the nova into the Fe\,II class of classical novae.

Our spectrum of V888\,Cen taken nine years after outburst shows faint 
emission lines for H$\alpha$, H$\beta$, H$\gamma$, as well as He\,II 
(469\,nm) and the Bowen blend at 460\,nm. 
The bluer Balmer lines are rather found in absorption.
This indicates a hot, optically thick accretion disc, typical for novalikes.
Since nine years is usually a sufficient time for a fast nova to cool down
\citep{schm+05}, 
we thus conclude that the binary V888\,Cen is a high accretion rate system.

\section{Summary}
We have discussed the spectroscopic classification for six candidate CVs. 
For two 
of them, LB\,9963 and FQ\,Mon, the CV classification could be confirmed, 
while the spectra of the others match those of normal
main sequence or evolved stars. 

LB\,9963 shows the 
typical emission lines embedded in absorption troughs, and is thus believed
to be a high mass transfer system. Arguing against this interpretation
is the lack of a blue continuum due to the disc. However, this
might be affected by high interstellar extinction. 

FQ\,Mon is strongly variable and has also been observed during outburst. 
However, the quiescence spectra show an inverse Balmer decrement, so we
do not believe this system to be a normal dwarf nova but rather a magnetic system.

We also
present the first spectrum of V888\,Cen in quiescence. We find
this old nova to 
be a high mass transfer system.
\acknowledgments
This material is based [in part] upon work supported by the National 
Science Foundation under Grant No. 0353843 in the 
framework of the CTIO-REU program. We also 
acknowledge that this research has made intense
use of the Simbad database operated at CDS, Strasbourg, France.



\begin{thebibliography}{}
\bibitem[Abrahamyan et al.(2003)]{abra+03}
Abrahamyan, G.V., Sinamyan, P.K., Gigoyan, K.S., 2003, Astrophysics, 46, 46

\bibitem[Berger \& Fringant(1980)]{berg+80}
Berger, J., Fringant, A.--M., 1980, \aaps, 39, 39

\bibitem[Downes et al.(2001)]{down+01} 
Downes R.A., Webbink R.F., Shara M.M., Ritter H., Kolb U., Duerbeck H.W., 2001,
\pasp, 113, 764, living edition

\bibitem[Drimmel et al.(2003)]{drim+03} 
Drimmel R., Cabrera-Lavers A., L\'opez-Corredoira M., 2003, \aap, 409, 205

\bibitem[Jones \& Stauffer(1991)]{jone+91}
Jones, B.F., Stauffer, J.R., 1991, \aj, 102, 1080

\bibitem[Hacke \& Richert(1990)]{hack+90}
Hacke, G., Richert, M., 1990, Ver\"off. d. Sternw. Sonnneberg, 10, 336

\bibitem[Hoffmeister(1936)]{hoffm36}
Hoffmeister, C., 1936, Astron. Nach., 259, 37

\bibitem[van Hoof(1941)]{hoof41}
van Hoof, A., 1941, Ciel et Terre, 57, 321

\bibitem[Horne(1986)]{horn86}
Horne, K. 1986, \pasp, 98, 609

\bibitem[Howarth(1983)]{howa83}
Howarth, I.D., 1983, \mnras, 203, 301

\bibitem[Kazarovets et al.(1999)]{kaza+99}
Kazarovets, A.V., Samus, N.N., Durlevich, O.V., Frolov, M.S., Antipin, S.V. 
Kireeva, N.N., Pastukhova, E.N., 1999, IBVS,  4659

\bibitem[Kilkenny(1995)]{kilk95}
Kilkenny, D., 1995, \mnras, 277, 920

\bibitem[Liller(1995)]{lille95}
Liller, W., 1995, \iaucirc, 6139

\bibitem[Masi(2004)]{masi04}
Masi, G., 2004, vsnet-superoutburst, 2297

\bibitem[Mermilliod et al.(1990)]{merm+90}
Mermilliod, J.-C., Weis, E.W., Duquennoy, A., Mayor, M., 1990, \aap, 235, 114

\bibitem[Munari \& Zwitter(1997)]{muna+97}
Munari, U., Zwitter, T, 1997, \aap, 318, 269

\bibitem[Petit(1960a)]{petit60a}
Petit, M., 1960a, AnAp, 23, 681

\bibitem[Petit(1960b)]{petit60b}
Petit, M., 1960b, Journ. Obs., 43, 17

\bibitem[Petit(1961)]{petit61}
Petit, M., 1961, Journ. Obs., 44, 6

\bibitem[Petersen \& Andreasen(1987)]{pete+87}
Petersen, J.O., Andreasen, G.K., 1987, \aap, 176, 183

\bibitem[Remillard et al.(1994)]{remi+94}
Remillard, R.A., Bradt, H.V., Brissenden, R.J.V., Buckley, D.A.H., 
Schwartz, D.A., Silber, A., Stroozas, B.A., Tuohy, I.R., 1994, \aap, 428, 785

\bibitem[Ritter \& Kolb(1998)]{ritt+98}
Ritter, H., Kolb, U., 1998, \aaps, 129, 83

\bibitem[Schlegel et al.(1998)]{schl+98}
Schlegel, D., Finkbeiner, D., Davis, M., 1998, \apj, 500, 525


\bibitem[Schmidtobreick et al.(2003)]{schm+03}
Schmidtobreick, L., Tappert, C., Bianchini, A., Mennickent, R.E., 2003, \aap,
410, 943

\bibitem[Schmidtobreick et al.(2005)]{schm+05}
Schmidtobreick, L., Tappert, C., Bianchini, A., Mennickent, R.E., 2005, \aap,
432, 199

\bibitem[Tappert \& Bianchini(2003)]{tapp+03}
Tappert, C., Bianchini, A., 2003, \aap, 401, 1101

\bibitem[Uemura(2004)]{uemur04}
Uemura, M., 2004, VSNET Weekly Campaign Summary, 1435

\bibitem[Warner(1995)]{warn95} 
Warner, B., 1995, Cataclysmic Variable Stars, Cambridge University Press

\bibitem[Yan Tse et al.(2001)]{yant+01}
Yan Tse, J., Hearnshaw, J.B., Rosenzweig, P., Guzman, E., Escalona, O., 
Gilmore, A.C., Kilmartin, P.M., Watson, L.C., 2001, \mnras, 324, 553
\end{thebibliography}
\end{document}